# Closing the AI generalization gap by adjusting for dermatology condition distribution differences across clinical settings

Rajeev V. Rikhye, PhD[1],  Aaron Loh, MS[1], Grace Eunhae Hong, BA[2], Preeti Singh, MS[1], Margaret Ann Smith, MBA[2], Vijaytha Muralidharan, MD[2], Doris Wong, BA[1], Rory Sayres, PhD[1], Michelle Phung, MS[2], Nicolas Betancourt, MD[2], Bradley Fong, BS[2], Rachna Sahasrabudhe BA[2], Khoban Nasim BS[2], Alec Eschholz, BA[2], Basil Mustafa, MEng[1], Jan Freyberg, PhD[1], Terry Spitz, MSc[1], Yossi Matias PhD[1], Greg S. Corrado PhD[1], Katherine Chou MS[1], Dale R. Webster PhD[1], Peggy Bui, MD, MBA[1], Yuan Liu, PhD[1], Yun Liu, PhD[1], Justin Ko, MD, MBA[2,*], Steven Lin, MD[2,*]

[1] Google Research, Mountain View, CA, USA

[2] Stanford University School of Medicine, Stanford, CA, USA

[*] These authors contributed equally

Address correspondence to:  stevenlin@stanford.edu, liuyun@google.com,

## Abstract

Recently, there has been great progress in the ability of artificial intelligence (AI) algorithms to classify dermatological conditions from clinical photographs. However, little is known about the robustness of these algorithms in real-world settings where several factors can lead to a loss of generalizability. Understanding and overcoming these limitations will permit the development of generalizable AI that can aid in the diagnosis of skin conditions across a variety of clinical settings. In this retrospective study, we demonstrate that differences in skin condition distribution, rather than in demographics or image capture mode are the main source of errors when an AI algorithm is evaluated on data from a previously unseen source. We demonstrate a series of steps to close this generalization gap, requiring progressively more information about the new source, ranging from the condition distribution to training data enriched for data less frequently seen during training. Our results also suggest comparable performance from end-to-end fine tuning versus fine tuning solely the classification layer on top of a frozen embedding model. Our approach can inform the adaptation of AI algorithms to new settings, based on the information and resources available.

## Abbreviations and Acronyms

CLIN (clinic-taken photographs); PAT (patient-taken photographs); AI (artificial intelligence)

# Introduction

Since the COVID-19 pandemic, skin conditions have grown to the fifth most common concern in telehealth in the United States[1]. As a result, more healthcare providers are now assessing dermatological conditions from digital photographs - such as those captured during clinical examination by a non-specialist or taken by the patient remotely. In parallel, several artificial intelligence (AI) algorithms have been developed to help interpret images of skin conditions[2]. While these algorithms can achieve close to or even better performance than clinicians in experimental settings[3–5], factors such as differences in case ambiguity[3,6,7], skin tone distribution[8] or dermatological condition differences between the training and evaluation set can lead to differences in AI algorithm accuracy and reduced usefulness in real-world settings. However, little is known about the robustness of these algorithms to previously unseen data sources[9], such as a different institution or data acquisition device.

In this retrospective study, we evaluated the performance of a deep learning model for skin condition classification on a dataset from Stanford Health Care, not previously seen by the model. Unlike both the training dataset and data used in most research studies, this dataset was enriched for patient-taken images (PAT, comprising approximately 25% of this dataset) instead of clinician-taken (CLIN, approximately 75%) images. Compared to the training set, there were substantial differences in the condition distribution, which we found contributed to lower accuracy of the model for this dataset. Here, we demonstrate a series of steps for refining the algorithm to help improve AI model generalization to new environments or dataset shifts.

# Methods

## Case selection

This study included 2,500 de-identified cases from Stanford Health Care's eConsult system between November 2015 and January 2021. The sample size was determined qualitatively with the goal of ensuring sufficient numbers of cases for analysis of CLIN/PAT differences and for subgroup analysis based on condition. This study was approved by the Institutional Review Board of Stanford University Medical Center (Protocol # 59290) and conducted according to the Declaration of Helsinki. To improve representation of rarer conditions, the data were downsampled *a priori* to retain 20% of cases with eczema and related conditions (ICD10 codes: L20-L30), 50% of cases with seborrheic keratoses/irritated seborrheic keratosis (SK/ISK) (ICD10 code: L82) and 50% of cases with rashes and nonspecific skin eruption (ICD10 code: R21). Unless otherwise indicated, all results were adjusted for this stratified sampling.

Age, sex, clinician-estimated Fitzpatrick skin type (eFST), photo source and other diagnostic metadata questions were extracted from the health record for each case. For cases with no FST in the health record, FST scores were visually assessed retrospectively. Cases with either "unknown" or "other" photo source were excluded (20/2,500). A small fraction of cases (79/2,500) had both CLIN and PAT images and were also excluded due to the ambiguity in the case-level PAT/CLIN classification and small numbers precluding robust analysis. CLIN cases were captured using the Epic HAIKU application on mobile phones. Cases containing video visit screenshots (14/2,500) were also excluded from this study to ensure that only photographs captured using a mobile phone camera are used.. The remaining cases were stratified into CLIN and PAT (**Figure 1A**). Note that information about the photographer for CLIN cases (e.g. MD, RN, or PA) was not available. For more details on the differences between CLIN and PAT, we refer the reader to **Supplementary Tables 1** and **2** and Rikhye et al.[10] .

**Case De-identification**

Case information, including images of skin conditions and clinical metadata, were stripped of all protected health information and subject identifiers. For each case, a Stanford research team member reviewed all images and clinical metadata. Any image containing identifying information (e.g., full face, unique traits such as a birthmark, piercing, or tattoo, etc.) was cropped or removed.

**Reference Diagnosis**

In addition to the Stanford dermatologist's assessment of the case, a panel of three US board-certified dermatologists independently provided a differential diagnosis and associated confidence score per diagnosis for each case. Each case included up to six photographs, demographic information and structured medical history. First, each of the three panel dermatologists assigned to a case provided a differential diagnosis and accompanying confidence values in the range [1-5] for each of the diagnoses. Next, each diagnosis was mapped to a list of 419 conditions. If duplicates occurred (i.e. multiple diagnoses were mapped to the same condition), the highest confidence was retained. Diagnoses that could not be mapped to the 419 conditions were labeled as "not categorized" and were subsequently dropped. Next, the conditions were ranked into a differential diagnosis based on their relative confidence values. Each condition in the differential diagnosis was then assigned a weight, which was the inverse of their rank. If multiple mapped conditions shared the same confidence, the weight was uniformly distributed across the conditions. For more details, we refer the reader to the supplementary methods in Liu et al[3].

Cases assessed by the panel dermatologists to have multiple conditions (19), undiagnosable conditions (12) due to image quality issues, minimal visible pathology, limited field-of-view, or conditions that were not supported were excluded from the study. Cases were then subclassified as unanimous (3/3 agreed), intermediate (2/3 agreed) or ambiguous (0/3 agreed) depending on the inter-rater agreements. Specifically, an easy case is one where all three panel dermatologists unanimously agreed on the diagnosis, while a hard case is one where none of the panel dermatologists agreed. A medium case, on the other hand, had two in agreement.

**Skin condition classification model**

The skin condition model used in this study (**Figure 1B**) resembles Liu et al.[3] with a few key differences. First, this model uses wide ResNet-101x3 feature extractors that were pretrained on web scale images using contrastive learning[11]. Briefly, each image had a paired text description of that image. Embeddings were computed for both image and text and a CLIP-style loss was minimized to ensure that similarity between similar conditions was maximized. For more detail, see Zhang et al.[12]

For each patient, we selected 6 images, at random for cases with more than 6 images. For each image, an image embedding was computed and embeddings across images were combined via the average-reduce operation. The metadata fields comprised 25 different categorical variables which included patient history questions (e.g. history of psoriasis, etc.), signs (e.g. redness, swelling, etc.), symptoms (e.g. itching, fever, etc.) and demographic factors (e.g. sex, gender and estimated skin type). All categorical questions were one-hot encoded and were combined with age before being projected into a metadata embedding.

The metadata embedding and image embedding were then combined using feature-wise linear modulation (FiLM) which is defined as follows:

$$FiLM(E_{image+metadata}) = \beta(E_{metadata}) + \alpha(E_{metadata}) \circ E_{image}$$

where, $\alpha(E_{metadata})$ and $\beta(E_{metadata})$ are learned projections of the metadata embeddings. Therefore, FiLM applies an affine transformation to the image embedding ($E_{image}$) by scaling it by $\alpha(E_{metadata})$ and shifting it by $\beta(E_{metadata})$. This in turn results in image embedding features that are modulated by features in the metadata. To make the model robust to missing metadata we used dropout where we replaced metadata fields with "unknown" values at random with a 25% probability.

To visualize the resulting embedding in two dimensions (**Figure 1C** and **Supplementary Figure 3**), we used UMAP to perform non-linear dimensionality reduction. The points, which represent each case, were then post-hoc colored by either skin condition category (**Supplementary Table 10**), anatomic location or photo source (CLIN, PAT).

The entire model was trained end-to-end on a training set of 41,290 cases (194,461 images) covering 204 unique primary conditions by minimizing the focal loss function[13] to account for class imbalance:

$$FL(p_t) = -\alpha(1 - p_t)^\gamma \ log(p_t)$$

where the parameters $\alpha$ and $\gamma$ are selected via hyperparameter tuning to maximize top-3 accuracy. Focal loss, which is a scaled form of cross entropy, was selected given the class imbalance in the development set, allowing the model to learn more from hard-to-classify examples. A constant learning rate schedule was used along with the ADAM optimizer and the number of training steps was also selected via hyperparameter tuning.

**Variable-k accuracy**

In addition to top-3 accuracy, where the top-1 ground truth has to be within the top-3 predictions, we also report top variable-k accuracy. Here, the number of predictions per case is customized on a case-by-case basis with smaller k for confident predictions and larger k for uncertain cases.

Using the development set, we computed a score threshold that maximized the sensitivity for high risk conditions (e.g. certain skin cancers) and minimized the false positive rate for an output between 3-7 predictions. During inference, we added prediction scores for the top-k predictions until the threshold was reached. Accuracy against the top-1 ground truth was then computed.

**Score recalibration**

The output of the model is a probability distribution over labels (419) and is said to be well calibrated if and only if $p(c_j \mid p^{c_j}) = p^{c_j}$, where $p^{c_j}$ is the probability estimate for class $j$. In other words, if a number of predictions with probability estimate 0.8 are made, then the predictions should be correct in 80% of the cases. To correct for the distribution differences between the DEV and Stanford data sets (Figure 3), we employed a multiclass version of Platt scaling[14]. Platt scaling fits a sigmoid function ($z$) to the scores ($z$) obtained by the model on the calibration set

$$\sigma(z^k) = exp(z^k/T)/ \sum_{j=1..K} exp(z^j/T)$$

Where T is the temperature parameter and k is the condition category. The parameter T is learned by performing one-vs-rest classification for each of the 12 condition categories on 20% of Stanford cases. This calibration set was identified by stratified sampling.

**Matching training data to Stanford condition distribution**

In order to match the DEV set to the Stanford condition distribution, we first compute the probability distribution of the Stanford dataset (CLIN and PAT combined) over the 29 condition categories. Next, we use the Metropolis-Hastings algorithm to generate samples from the DEV set with a fixed random seed. The AI was either retrained or fine-tuned using this DEV dataset.

**Augmenting training data with Stanford examples**

In order to augment the DEV set with examples from Stanford, we first identified conditions in the Stanford dataset that were less common in the DEV set by comparing the two histograms. Next, we sampled cases from the less common conditions with probability $\alpha = 0.7$ and cases from more common conditions with probability $1 - \alpha$. These cases were then added to the existing training set. In this way, the model was trained on both Stanford conditions and images of those conditions. The sampled cases were drawn from a 20% subset of PAT and CLIN (stratified by condition and demographics), and evaluation was done using the remaining 80%. To ensure comparability across these experiments (recalibration, training data changes, tuning), the evaluations used the same 80% set (**Supplementary Table 7**).

# Results

The deep learning model (**Figure 1B**) used in this study takes as input between 1 to 6 photographs and a list of 25 structured metadata questions (such as medical history information) and returns a prediction over 419 possible skin conditions. The architecture of the model is an improvement upon the one developed by Liu et al.[3] in a few key ways. First, the image embeddings for each digital photograph are computed using a pretrained ResNet-101[15] architecture pretrained using large-scale web data[16,17]. Second, feature wise linear modulation (FiLM)[18,19] was used to fuse the metadata embeddings with the image embedding. This method preserves the embedding size and modulates image features with metadata features, leading to better results empirically.

We visualized the resulting embedding by projecting it to two dimensions using UMAP [20] (**Figure 1C**) and found that the embedding clusters by condition category and body location. For example, eruptions on the face are distinct (in embedding space) to eruptions on the torso. Interestingly, we do not notice clustering by photograph source (CLIN vs. PAT), which suggests that the AI is able to generalize across this aspect and extracts the same features regardless of who took the photograph (i.e., clinically trained vs. untrained individuals).

We first sought to understand if the AI which was trained primarily (though not exclusively) on clinician taken images (CLIN) could generalize to patient-taken images (PAT), which are, on average, poorer quality than photographs taken by clinicians in prior studies[21,22]. We measured performance in terms of top-3 accuracy where the three highest scoring model predictions must contain the top reference diagnosis. Surprisingly, we found no difference in the AI's top-3 accuracy between CLIN and PAT (**Table 1**, top-3 accuracies between 71% and 74%), suggesting that the AI performed similarly on both photo sources. Comparisons with an internal test dataset (i.e. cases from similar sources to that used in training, but where the specific cases were not used for training) are shown in **Supplementary Table 3**.

Interestingly, Stanford dermatologists' (Derm) diagnoses were more in agreement with the reference diagnosis for PAT than for CLIN (**Table 1**). This higher CLIN accuracy resulted in a greater difference between Derm and AI performance for PAT (17%) than for Clin (5%). In other words, even though AI performance was similar between the two sets, AI performance relative to dermatologists differed, suggesting that the AI performance was lower after "controlling" for dermatologist performance. **Supplementary Table 4** shows the top-3 accuracy stratified by case ambiguity, defined using the inter-dermatologist agreement rate among the 3 dermatologists who retrospectively reviewed each case. As expected, both the AI and Derm have higher accuracy for easier (where all 3 dermatologists are in agreement) cases than the hard cases (without inter-dermatologist agreement). For the least ambiguous cases, the AI performs better on CLIN than PAT (82% and 72% respectively), while Derms have comparable accuracy (>90%).

Increasing the number of conditions output by the model improves the top-k accuracy, since there is a higher likelihood of including the reference diagnosis. However, this approach also leads to a higher false positive rate by outputting less relevant conditions. We developed a method[23] that allowed us to dynamically select the $k$ diagnoses on a case-by-case basis while maintaining the overall sensitivity at 95%. With this advance, using top-k accuracy, we found similar performance between the AI and Derm for both CLIN and PAT (**Table 1**;  83-85% for the AI vs. 81-89% for Derm). We also observed no differences in the number of conditions output by the model between CLIN and PAT cases (**Supplementary Table 5**). This suggests that the model had identified relevant conditions (e.g. within the top 7), but was not ranking them highly enough (e.g. within the top 3) to achieve comparable top-3 accuracies.

### *Demographic and image quality factors with model performance*

To what extent are differences in demographic (**Supplementary Table 1**), clinical (**Supplementary Table 2**), and image quality factors[10] responsible for the differences in model performance for CLIN and PAT images? We addressed this question by performing multivariable logistic regression (see **Methods**) to determine the factors correlated with either correct AI or Derm predictions. The forest plot in **Figure 2**, shows that demographic factors such as age, sex and estimated Fitzpatrick Skin Type (eFST) were not associated (p>0.05) with either the AI or the Derm making a correct prediction (see also **Supplementary Table 6**). This further suggests that the AI accuracy (and errors) are not associated with the demographic information above. Relative to ambiguous cases, where none of the three dermatologists on the panel agreed on the top diagnosis, unambiguous and intermediate (2 of 3 labeling dermatologists in agreement) cases were associated (p=0.00057 and p=0.00038 respectively) with correct predictions for both the AI and Derm. This further underscores the finding that case ambiguity or difficulty has an important impact on both AI and Derm accuracy.

Similar to Derm, the AI was fairly robust to image quality differences and only images with significant non-skin areas, such as clothes or background, were associated (p<0.05) with incorrect AI predictions. The effect was similar in direction for Derms but was not statistically significant (p=0.28). Interestingly, while the skin condition category had no impact on Derm accuracy, the AI made more correct predictions for certain categories, such as neoplasms, inflammatory conditions and hair conditions. Anatomic location of the skin location has no influence on model performance, though lesions on genitalia were associated with lower Derm accuracy. Together this suggests that AI accuracy could be impacted by the presence of conditions that the model is not expecting. Interestingly, a trend (albeit non-significant post adjustment) was observed where Derm was more accurate in later years (the reverse of that seen for the AI), possibly signaling a learning effect by dermatologists getting used to the eConsult system, or a shift towards greater utilization and easier condition distributions with time.

### *Condition distribution differences contribute to lower model performance*

As highlighted by the multivariable regression analysis, certain skin conditions categories are significantly associated with AI model performance. We hypothesized that differences in condition distributions between the AI development dataset and the Stanford dataset may be responsible for relative AI-Derm differences in performance across PAT and CLIN cases. In particular, inflammatory conditions (such as acne, eczema and psoriasis), neoplasms (melanocytic nevus) and contact dermatitis were much more prevalent in the AI development dataset, while cutaneous infections (such as intertrigo, tinea, herpes zoster) were more prevalent in both Stanford subsets. The cases for which the model was more confident (i.e. max classifier score > threshold) had condition distributions that were more similar to the AI development dataset compared to lower confidence cases. Similarly, the long tail of conditions (in the AI development dataset) had a higher incidence in the low confidence cases (**Figures 3A-B**).

To explore this notion further, we sub-sampled cases in both CLIN and PAT datasets to match the DEV set distribution (**Figures 3C-D**). Evaluating the AI on these cases resulted in an 10.0% absolute improvement in accuracy for CLIN and 8.4% for PAT (**Table 2**). Altogether, these results suggest that dermatological AI model generalization to new datasets is hindered by an unexpected condition distribution. In particular, the closer the distribution is to the training set, the better will be the model performance.

### *Fine tuning strategies for closing this generalization gap*

We next explored four different methods for closing the generalization gap, using progressively more information about the CLIN and PAT datasets, and more computational resources (**Figure 4** and **Supplementary Tables 7-9**).

First, we developed a multiclass version of Platt scaling[14] to recalibrate the scores to the distribution expected in the Stanford dataset. Briefly, we held out 20% of the Stanford dataset (via stratified sampling to ensure identical distributions) and learned condition-specific temperature parameters that recalibrated the softmax scores such that the predicted distribution more closely matched the new distribution (**Figure 3A**). This resulted in a 4% and 8% absolute improvement in top-3 accuracy for CLIN and PAT respectively (**Figure 3C**). Despite this, a gap remained between the AI and the Derm.

Second, similar in spirit to the test set condition distribution resampling experiment in **Figure 3**, we resampled the training distribution to match the Stanford distribution and retrained the entire model end-to-end with this new DEV dataset. Since this resampling was done in a condition and demographics-aware manner, we found no significant difference in the demographic distribution of the resampled evaluation dataset (see **Methods** and **Supplementary Table 7**). Score recalibration was performed on the re-trained model as mentioned above. This further (albeit to a smaller extent) improved the model performance, underscoring the importance of condition distribution matching in AI generalization. It is important to note that in both these methods, the model is not exposed to patient images from Stanford and any improvements are solely through using information about the different condition distribution.

Third, we developed a condition-aware sampling method that augmented the existing DEV set with "unexpected" condition cases (i.e., those less common in DEV) from the Stanford dataset. Specifically, we sampled conditions that are less prevalent in the training set with a higher probability (0.7) while sampling more prevalent conditions with a lower probability (see **Figure 3B**). Unlike the previous methods, this augments the training set with images from Stanford. We retrained the entire model end-to-end and found a ~13% improvement in top-3 accuracy for both CLIN and PAT sets. This condition-aware sampling method was better than the common approach of augmenting the training set by randomly selecting 20% of the data (**Supplementary Table 8**). Notably, this method of fine tuning did not change the equity of the model across demographic factors (**Supplementary Table 9**).

Finally, we noticed that following retraining, the post-FiLM embeddings do not change substantially (**Supplementary Figure 1**). Therefore we hypothesized that a computationally more efficient approach would be to freeze both the image and metadata encoders and fine-tune only the final classification layer. Relative to re-training this model, this method also improved top-3 accuracy for CLIN and PAT by ~13% (**Figure 4C**).

Taken together, our methods suggest that an efficient way to adjust a dermatological AI model to a new clinical setting with different condition distribution is to (1) augment the existing training data with conditions that are unexpected or less frequent and (2) fine tune the classifier layer along with a temperature-based score recalibration to adjust the model confidence to match the new expected distribution.

## Discussion

In recent years, there has been a growing interest in the development of AI technologies that can interpret skin conditions from digital photographs[2]. These and other similar technologies have great promise of augmenting the clinical decision making process[24–26]. However, most of these have been tested only in experimental settings and it is currently unknown how the AI can generalize to data from different clinical settings. We retrospectively assessed the generalizability of an AI model for skin condition detection on a dataset not previously seen by the model of 2,500 eConsult cases from Stanford Health Care with different photo sources and a different condition distribution to the development set. The AI model generalized well to both clinician-taken and patient-taken photographs and its accuracy was not associated with any demographic factors. Case ambiguity and condition distribution differences relative to the development set were the factors most strongly associated with cases with poorer accuracy.

Experimenting with different fine-tuning methods, we found that the best way to close the generalization gap is to augment the development set with unexpected or less frequent conditions and to fine tune the final classification layers of the AI. This method requires relatively low computational resources as it does not require retraining of the image or metadata encoders. The comparability of fine tuning solely the classification layer versus end-to-end fine tuning bodes well for generalizing models developed using frozen foundation models where the encoder is frozen and one or more layers are trained on top of the frozen model. In terms of augmentation, in the absence of labeled data from the target distribution, some of these improvements can be obtained using solely *a priori* information about the anticipated condition distribution in the target clinical setting. These could include anticipated seasonal changes or known geographical differences in disease distribution relative to the training set.

The AI model that we introduce in this paper has several unique features. First, the feature extractors were pre-trained on unlabelled web-scale data and fine-tuned on a mixture of clinical photographs, dermatoscope images and user-taken photographs. Metadata was fused with the image embedding using a technique called feature-wise linear modulation (FiLM), which preserves the size of the embedding, permits smaller models[19] and learns to map both metadata and image features to the same space[27]. As we have demonstrated from fine-tuning results, these techniques make the combined image-metadata embedding highly informative. In other words, by grouping into body location and condition category, the combined embedding encodes the information required for interpreting many dermatologic conditions. Interestingly, the embedding does not cluster by photograph source, which implies that representation of features from CLIN or PAT photographs are similar. In other words, features of a rash on a patient's face from different photographers, devices, and settings are more similar than that of rashes on different anatomic locations.

Our first key finding is that the AI model performs equally well for both CLIN and PAT photographs (when measured using top-3 accuracy). Teledermatology, where the patient either sends photographs asynchronously or consults with their care provider via video, is becoming increasingly popular[28,29]. As a result, dermatologists have to review both patient-taken and clinician-taken photographs of skin conditions. Our result is meaningful because it suggests that an AI model can be deployed to a system that includes both photograph sources without requiring substantial fine tuning or having separate models for each photograph source. We show that dermatologists tend to perform better on the PAT dataset, and observe that differences can be attributed to case ambiguity because predictions for ambiguous cases tend to be less accurate. For ambiguous cases – i.e. those where the differential diagnosis has uncertainties based on the images and structured medical history alone – dermatologists tend to rely on other information and tests, such as additional examination and targeted questions or biopsies to narrow down the differential diagnosis[30]. One limitation of our study is that we do not have information about recommended next steps, such as biopsy or in-clinic follow-up. This information would allow us to better understand the source of ambiguity in these cases.

Our second key finding is that condition distribution differences contribute to the generalization gap seen when the AI model was evaluated on the previously unseen dataset. Traditionally, most AI algorithms are trained based on the assumption that both the training and testing data are identically and independently distributed[31]. However this assumption rarely holds in reality, and when training and testing probability distributions are different, prediction accuracy can deteriorate[32] as we have demonstrated. Accounting for condition distribution differences is important because skin diseases vary widely by time (including both seasonal and non-seasonal variations) and geographic location[33,34] . Our proposed method of adjusting the training or prediction distribution to match the expected condition distribution and fine tuning the classifier can help mitigate AI performance regressions in these scenarios. One limitation of the training adjustment method, however, is that it requires labeled data which may be costly or challenging to obtain in the setting of rapid shifts. Alternatives include the prediction distribution recalibration based on expected changes and targeted labeling based on data point mapping in the embedding space, which appear relatively invariant to distribution shifts.

In sum, our study presents a set of techniques based on the amount of information and data available to fine tune dermatological AI models to condition distribution differences that can arise in different clinical settings. We demonstrate that by adjusting the model to match an expected condition distribution, our AI achieves dermatologist-level accuracy on a previously unseen dataset.

# Figures

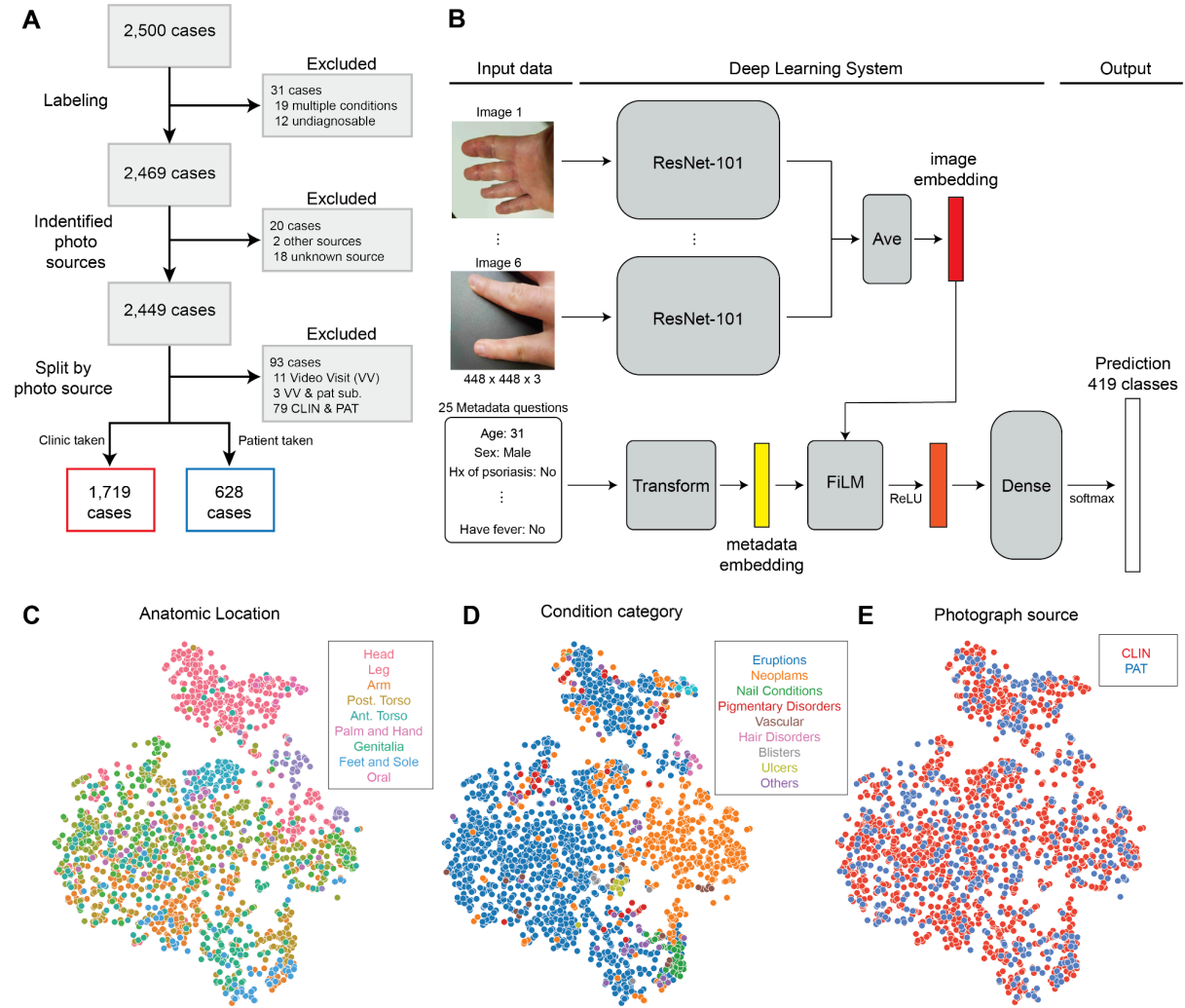

**Figure 1: Study and model overview and visualization of case embeddings.** A: Case selection procedure. Characteristics of the PAT and CLIN datasets are presented in Supplementary Tables 1 and 2. B: Model architecture. C-E: Two dimensional projections of the combined image-metadata embedding colored by anatomic location (C), condition category (D) and photograph source (E).

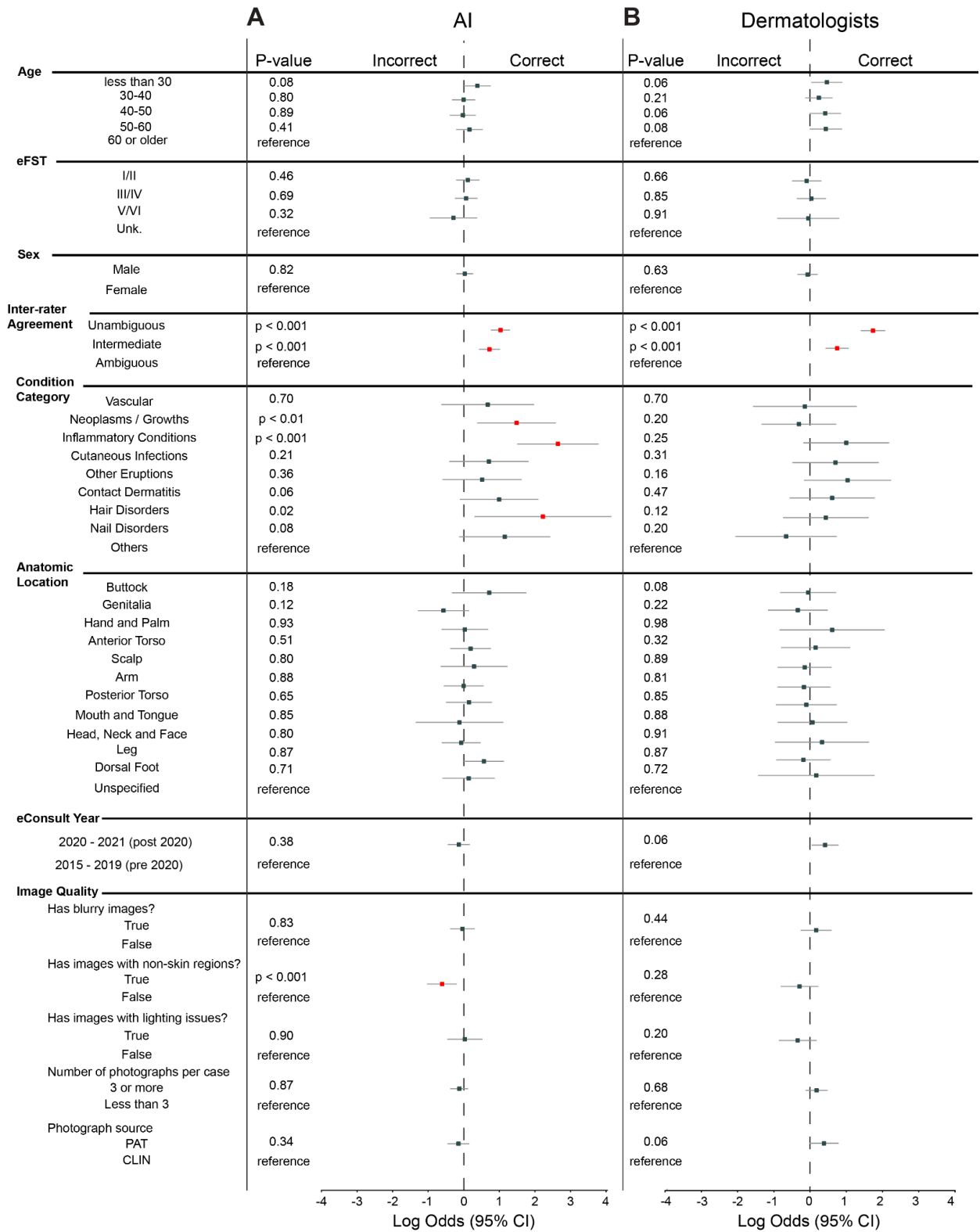

**Figure 2: Association between demographic, clinical and image quality factors and accuracy.** Within each group, points are sorted in descending order based on their log odds. Red

colored points are statistically significant at an alpha level of 0.05 after Bonferroni correction per variable (e.g., 3 age groups). The asterisks indicate that the point is outside the range of the graph, and indicated separately to aid visualization of the remaining points. Log odds were used instead of odds ratio to ensure left-right visual symmetry. Both the AI and dermatologists are less accurate in more difficult cases with inter-rater disagreement within the dermatologist panel. For the AI, several condition categories and images with substantial amounts of non-skin areas were also significantly associated with errors.

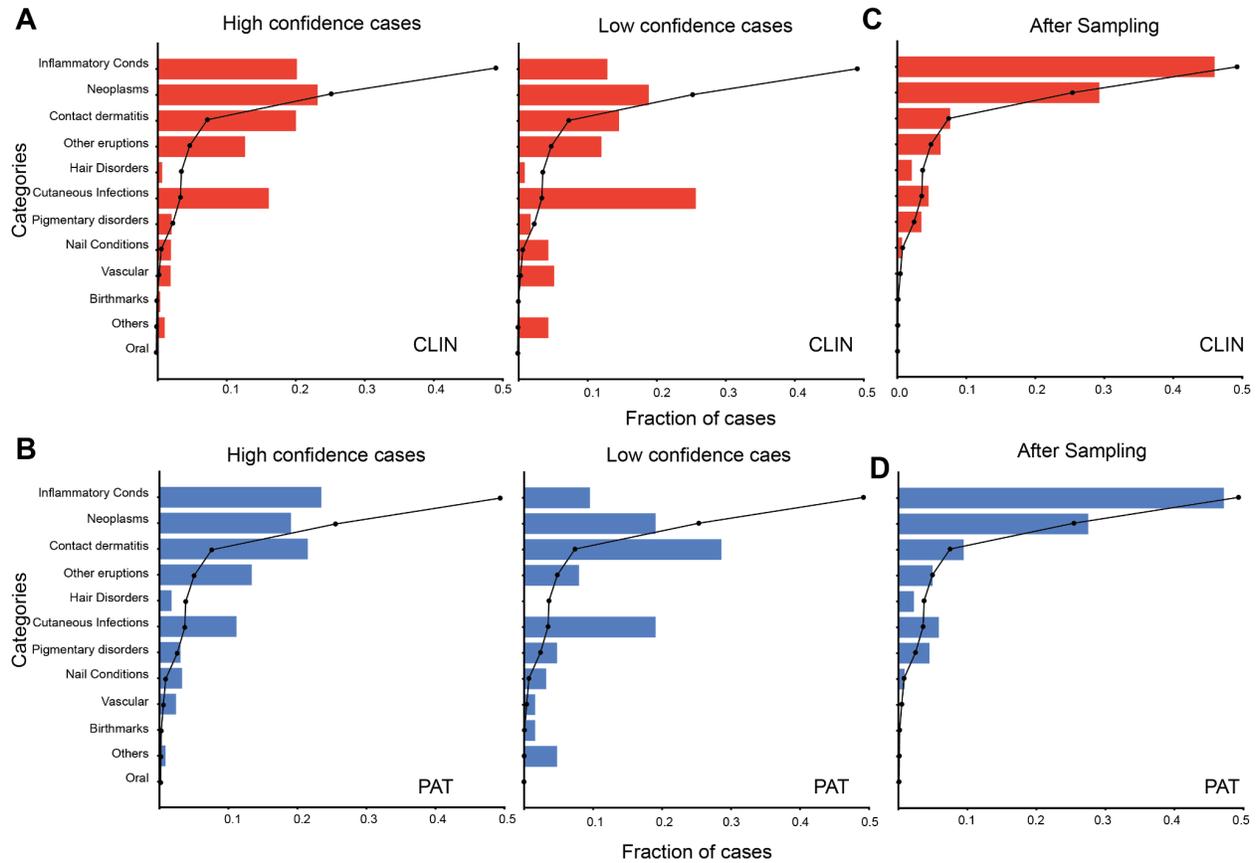

**Figure 3: Comparison of condition distributions.** A-B: Comparison between CLIN (A, red) and PAT (B, blue) relative to the DEV set (black) for both high (left) and low (right) confidence cases. C-D: Condition distributions for CLIN and PAT after resampling to match the DEV set distribution.

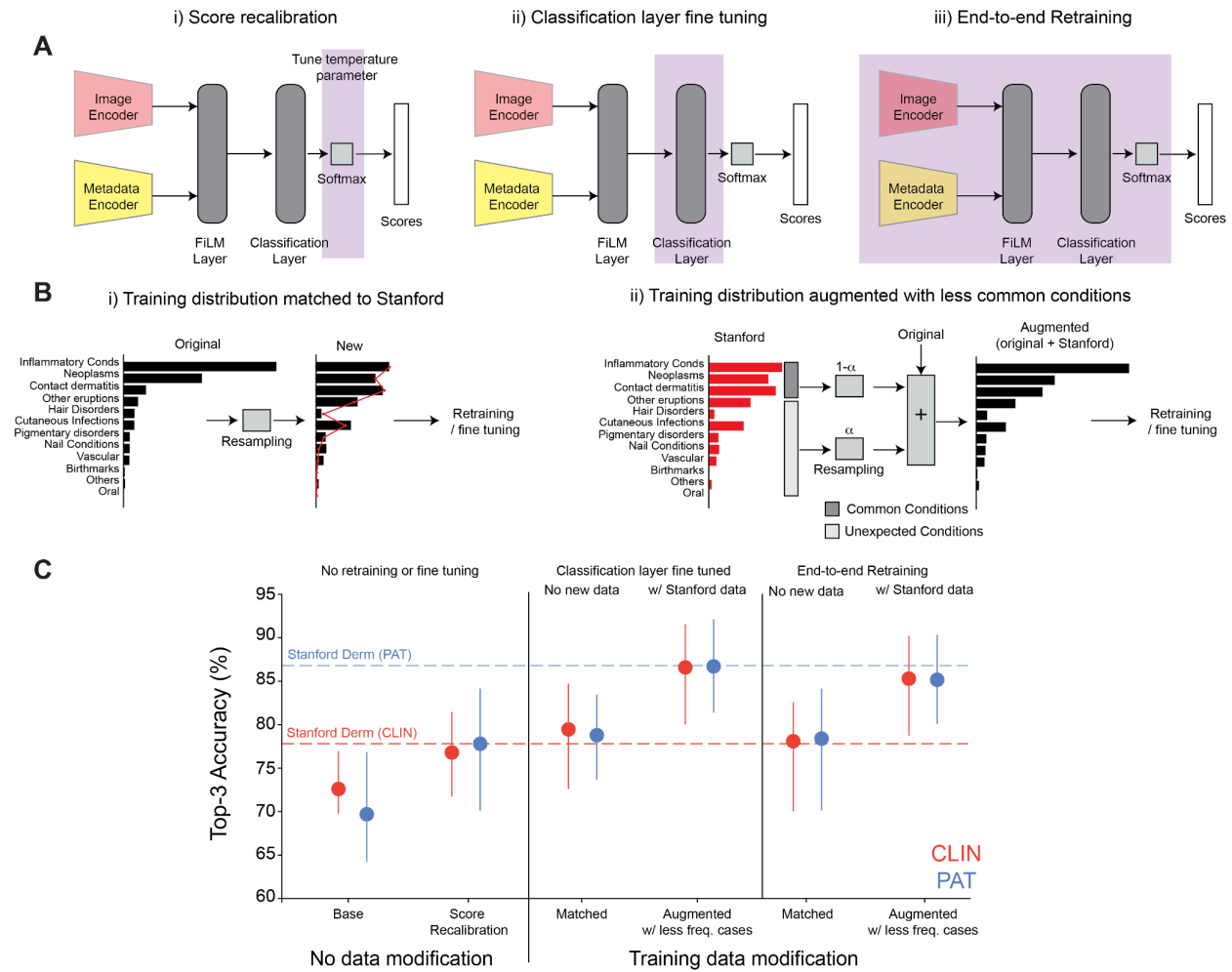

**Figure 4: Fine tuning closes the gap between AI and Stanford Dermatologists. A:** Schematic illustrating the various model changes made. **B:** Schematic illustrating the two different modifications applied to the training data. In (i) the training distribution is resampled to match the Stanford distribution. In (ii) the training distribution is augmented with case examples from Stanford of less frequent conditions. **C:** Summary plot showing changes in top-3 accuracy after applying different fine tuning methods. Error bars represent 95% confidence intervals.

# Tables

**Table 1: Comparison between AI and Stanford dermatologist (Derm) accuracy for both CLIN and PAT.** The 95% confidence intervals are shown in parentheses.

| | CLIN (n = 1,681 patients) | | PAT (n = 587 patients) | |
|---|---|---|---|---|
| | **AI** | **Stanford Derm** | **AI** | **Stanford Derm** |
| **Top-3 Accuracy** | 73.6% (71.2, 75.9) | 78.7% (76.4, 80.9) | 70.7% (66.3, 74.8) | 87.3% (83.7, 90.3) |
| **Top-k Accuracy** | 84.6% (82.8, 86.4) | 80.5% (78.2, 82.6) | 82.5% (78.9, 85.7) | 88.7% (85.4, 91.6) |

**Table 2: Comparison between AI and Stanford dermatologist (derm) accuracy for both CLIN and PAT after resampling 500 cases each to match the DEV distribution.**

|  | Re-sampled CLIN (n = 500 patients) | | Re-sampled PAT (n = 500 patients) | |
|---|---|---|---|---|
|  | **AI** | **Stanford Derm** | **AI** | **Stanford Derm** |
| **Top-3 Accuracy** | 83.6% (78.7, 87.8) | 77.3% (71.1, 82.7) | 79.1% (72.9, 84.5) | 88.9% (83.5, 93.0) |
| **Top-k Accuracy** | 93.0% (89.4, 95.6) | 79.5% (73.6, 84.5) | 87.8% (82.8, 91.8) | 91.9% (86.7, 95.4) |

# Supplementary Material

**Supplementary Table 1: Descriptive statistics of demographic variables in this dataset.**

|  |  | **CLIN** | **PAT** |
|---|---|---|---|
| **Total number of cases** | | 1,681 | 587 |
| **Sex (n, %)** | Female | 807 (48.0%) | 320 (59.3%) |
| | Male | 670 (39.9%) | 219 (40.6%) |
| **Age (n, %)** | ≥ 60 | 409 (24.3%) | 125 (23.2%) |
| | 50-59 | 210 (12.5%) | 79 (14.7%) |
| | 40-49 | 218 (13.0%) | 97 (18.0%) |
| | 30-39 | 355 (21.1%) | 145 (26.9%) |
| | < 30 | 285 (17.0%) | 93 (17.3%) |
| **eFST (n, % excluding unknown)** | Unk. | 289 | 247 |
| | V/VI | 53 (4.5%) | 10 (3.42%) |
| | III/IV | 548 (46.1%) | 145 (49.7%) |
| | I/II | 587 (49.4%) | 137 (46.9%) |

**Supplementary Table 2: Descriptive statistics of clinical variables in this dataset.**

| | | CLIN pre-2020 | PAT |
|---|---|---|---|
| **Condition Category (n, %)** | Infections | 240 (14.3%) | 63 (11.7%) |
| | Contact dermatitis | 271 (23.2%) | 113 (21.0%) |
| | Inflammatory | 280 (24.0%) | 112 (20.8%) |
| | Other eruptions | 183 (15.7%) | 65 (12.1%) |
| | Neoplasms | 334 (28.6%) | 97 (18.0%) |
| | Pigmentary disorder | 34 (2.9%) | 17 (3.0%) |
| | Vascular | 29 (2.5%) | 12 (2.2%) |
| | Blisters and ulcers | 16 (1.4%) | 5 (0.9%) |
| | Hair disorders | 10 (0.9%) | 8 (1.5%) |
| | Nail disorders | 30 (2.6%) | 17 (3.2%) |
| | Others | 49 (4.2%) | 30 (5.6%) |
| **Anatomic Location (n, %)** | Scalp | 32 (2.7%) | 18 (3.34%) |
| | Face and neck | 346 (29.6%) | 132 (24.5%) |
| | Mouth and tongue | 15 (1.3%) | 4 (0.7%) |
| | Arm | 232 (19.9%) | 73 (13.5%) |
| | Hand and palm | 87 (7.4%) | 28 (5.2%) |
| | Anterior Torso | 191 (16.4%) | 72 (13.4%) |
| | Posterior Torso | 103 (8.8%) | 37 (6.9%) |
| | Genitalia | 61 (5.2%) | 22 (4.1%) |
| | Buttocks | 31 (2.7%) | 5 (0.9%) |
| | Leg | 254 (21.7%) | 100 (18.6%) |
| | Dorsal Foot | 40 (3.4%) | 26 (4.8%) |
| | Unspecified | 85 (7.3%) | 22 (4.1%) |
| **Inter-rater agreements** | Disagreement (0/3) | 459 (39.2%) | 145 (26.9%) |
| | Intermediate (2/3) | 342 (29.3%) | 145 (26.9%) |
| | Unanimous (3/3) | 646 (55.3%) | 249 (46.2%) |

**Supplementary Table 3: Comparison of top-3 accuracy between the held-out DEV set and Stanford datasets**. DEV-matched refers to DEV cases that were resampled to match the Stanford distribution. All n values refer to number of patients.

|  | **CLIN (n = 1,681)** | **PAT (n = 569)** | **DEV (n=1000)** | **DEV Matched (n = 1000)** |
|---|---|---|---|---|
| **AI** | 72.6% (69.8, 76.9) | 69.7% (64.3, 76.8) | 82.6% (81.9, 91.6) | 78.9% (76.2, 81.9) |

**Supplementary Table 4: Top-3 Accuracy stratified by case ambiguity as measured by inter-rater agreement.**

| Dermatologist panel agreement rate | CLIN | | | PAT | | |
|---|---|---|---|---|---|---|
| | **% of data (n patients)** | **AI** | **Derm** | **% of data (n patients)** | **AI** | **Derm** |
| Unanimous agreement ("**easy**") | 29.5% (557) | 0.8156 [0.777, 0.849] | 0.915 [0.887, 0.937] | 35.1% (206) | 0.7216 [0.649, 0.786] | 0.949 [0.9087, 0.975] |
| 2 of 3 agreed ("**medium**") | 29.5% (470) | 0.7457 [0.703, 0.785] | 0.746 [0.702, 0.786] | 29.5% (154) | 0.7568 [0.679, 0.823] | 0.873 [0.807, 0.923] |
| Complete disagreement ("**hard**") | 40.7% (284) | 0.6299 [0.591, 0.668] | 0.616 [0.572, 0.659] | 35.3% (65) | 0.6158 [0.539, 0.688] | 0.729 [0.645, 0.803] |

**Supplementary Table 5**: **Number of conditions predicted by the model (K) distribution for CLIN and PAT.** There is no significant difference observed between CLIN and PAT.

|  | CLIN<br>(n = 1,681 patients) | PAT<br>(n = 587 patients) |
|---|---|---|
| **Average k<br>[IQR]** | 6.187<br>[3, 7] | 6.233<br>[3, 7] |

**Supplementary Table 6: Top-3 Accuracy stratified by demographic factors.**

| Variable | Category | CLIN | | PAT | |
|---|---|---|---|---|---|
| | | AI | Derm | AI | Derm |
| Sex | Female | 0.663 [0.631, 0.693] | 0.767 [0.736, 0.796] | 0.673 [0.619, 0.724] | 0.875 [0.832, 0.911] |
| | Male | 0.672 [0.636, 0.705] | 0.769345 [0.736, 0.801] | 0.677 [0.611, 0.739] | 0.851 [0.793, 0.898] |
| Age | < 30 | 0.765 [0.713, 0.811] | 0.803 [0.752, 0.847] | 0.682 [0.582, 0.774] | 0.893 [0.807, 0.946] |
| | [30, 40) | 0.625 [0.576, 0.672] | 0.759 [0.712, 0.801] | 0.606 [0.505, 0.700] | 0.875 [0.805, 0.927] |
| | [40, 50) | 0.663 [0.601, 0.721] | 0.799 [0.741, 0.849] | 0.606 [0.505, 0.700] | 0.888 [0.808, 0.943] |
| | [50, 60) | 0.646 [0.581, 0.706] | 0.762 [0.698, 0.818] | 0.671 [0.556, 0.773] | 0.871 [0.770, 0.939] |
| | >= 60 | 0.648 [0.601, 0.692] | 0.737 [0.690, 0.780] | 0.694 [0.604, 0.775] | 0.806 [0.714, 0.879] |
| eFST | I-II | 0.66755 [0.633, 0.701] | 0.745 [0.711, 0.778] | 0.660819 [0.585, 0.731] | 0.901 [0.841, 0.943] |
| | III-IV | 0.656 [0.618, 0.692] | 0.777 [0.741, 0.810] | 0.667 [0.590, 0.737] | 0.836 [0.767, 0.891] |
| | V-VI | 0.623 [0.490, 0.744] | 0.800 [0.670, 0.896] | 0.750 [0.349, 0.968] | 0.714 [0.290, 0.963] |
| | Unknown | 0.708995 [0.639, 0.773] | 0.820988 [0.753, 0.877] | 0.692 [0.622, 0.756] | 0.869 [0.809, 0.915] |

**Supplementary Table 7: Descriptive statistics of demographic variables dataset used to evaluate the fine-tuned model.** Compared with Supplementary Table 1, there is no significant difference in the distribution of demographic variables in the resampled dataset.

| | | CLIN | PAT |
|---|---|---|---|
| **Total number of cases after resampling** | | 1,340 | 400 |
| **Sex (n, %)** | Female | 777 (58.0%) | 236 (59%) |
| | Male | 563 (42.0%) | 164 (41%) |
| **Age (n, %)** | ≥ 60 | 335 (25.0%) | 112 (28.0%) |
| | 50-59 | 241 (18.0%) | 68 (17.0%) |
| | 40-49 | 146 (11.0%) | 84 (21.0%) |
| | 30-39 | 268 (20.0%) | 88 (22.0%) |
| | < 30 | 348 (26.0%) | 48 (12.0%) |
| **eFST (n, % excluding unknown)** | Unk. | 150 | 100 |
| | V/VI | 60 (5.0%) | 9 (3.0%) |
| | III/IV | 562 (47.2%) | 153 (51.0%) |
| | I/II | 641 (47.8%) | 138 (46.0%) |

**Supplementary Table 8: Comparison between AI and dermatologist accuracy for both CLIN and PAT on a hold out set after two different fine tuning strategies were used to re-train the model.** The 95% confidence intervals are shown in parentheses.

| | | CLIN (n = 1,340 patients) | | PAT (n = 400 patients) | |
|---|---|---|---|---|---|
| | | AI | Derm | AI | Derm |
| **Baseline (on this subset)** | | 72.6% (69.8, 76.9) | 77.8% (75.3, 81.2) | 69.7% (64.3, 76.8) | 86.8% (82.7, 89.4) |
| **Score recalibration** | Same model | 76.8% (71.8, 81.4) | | 77.8 % (70.2, 84.1) | |
| **Training distribution matched to Stanford** | Model **retrained** w/o Stanford case images | 78.1% (70.1, 80.2) | | 78.4 % (70.2, 84.1) | |
| **Random split (20%)** | Model **retrained** w/ Stanford case images | 74.2% (69.9, 78.1) | | 77.1% (70.2, 83.0) | |
| **Condition aware split (20%)** | | 83.7% (78.2, 89.1) | | 80.2% (76.9, 87.5) | |
| **Condition aware split (20%) + score recalibration** | | 85.3% (78.8, 90.2) | | 83.1% (78.1, 88.2) | |
| **Condition aware split (20%)** | Classifier **Fine tuned** w/ Stanford case images | 84.8% (78.8, 89.6) | | 83.4% (77.9, 88.5) | |
| **Condition aware split (20%) + score recalibration** | | 85.9% (79.1, 91.1) | | 84.2% (79.1, 88.8) | |

**Supplementary Table 9: Top-3 Accuracy stratified by demographic factors before and after fine tuning the model.**

| Variable | Category | CLIN | | PAT | |
|---|---|---|---|---|---|
| | | **Before** | **After** | **Before** | **After** |
| Sex | Female | 0.663 [0.631, 0.693] | 0.787 [0.741, 0.796] | 0.673 [0.619, 0.724] | 0.857 [0.821, 0.911] |
| | Male | 0.672 [0.636, 0.705] | 0.780 [0.742, 0.801] | 0.677 [0.611, 0.739] | 0.871 [0.789, 0.898] |
| Age | < 30 | 0.765 [0.713, 0.811] | 0.813 [0.732, 0.847] | 0.682 [0.582, 0.774] | 0.839 [0.817, 0.946] |
| | [30, 40) | 0.625 [0.576, 0.672] | 0.779 [0.722, 0.801] | 0.606 [0.505, 0.700] | 0.851 [0.815, 0.927] |
| | [40, 50) | 0.663 [0.601, 0.721] | 0.799 [0.761, 0.849] | 0.606 [0.505, 0.700] | 0.899 [0.808, 0.943] |
| | [50, 60) | 0.646 [0.581, 0.706] | 0.778 [0.687, 0.818] | 0.671 [0.556, 0.773] | 0.877 [0.780, 0.939] |
| | >= 60 | 0.648 [0.601, 0.692] | 0.747 [0.691, 0.780] | 0.694 [0.604, 0.775] | 0.816 [0.714, 0.879] |
| eFST | I-II | 0.66755 [0.633, 0.701] | 0.755 [0.701, 0.778] | 0.660819 [0.585, 0.731] | 0.911 [0.841, 0.943] |
| | III-IV | 0.656 [0.618, 0.692] | 0.777 [0.711, 0.810] | 0.667 [0.590, 0.737] | 0.863 [0.767, 0.891] |
| | V-VI | 0.623 [0.490, 0.744] | 0.810 [0.671, 0.896] | 0.750 [0.349, 0.968] | 0.741 [0.390, 0.963] |
| | Unknown | 0.708995 [0.639, 0.773] | 0.821 [0.754, 0.877] | 0.692 [0.622, 0.756] | 0.888 [0.819, 0.915] |

## Supplementary Table 10: Categorization of skin conditions.

| Category | Conditions |
|----------|-----------|
| Contact dermatitis | Allergic Contact Dermatitis, Irritant Contact Dermatitis |
| Cutaneous Infections | Impetigo, Abscess, Cellulitis, Syphilis, Ecthyma, Erythrasma, Pitted keratolysis, Tinea, Deep fungal infection, Herpes Zoster, Herpes Simplex, Molluscum Contagiosum, Hand foot and mouth disease, Chicken pox exanthem, Insect Bite, Tinea Versicolor, Scabies, Paronychia, Candida, Skin and soft tissue atypical mycobacterial infection |
| Inflammatory Eruptions | Eczema, Psoriasis, Acne, Rosacea, Perioral Dermatitis, Intertrigo, Hidradenitis, Perleche, Folliculitis, Lichen planus/lichenoid eruption, Seborrheic Dermatitis, Chilblain, Pigmented purpuric eruption, Granuloma annulare, Infected eczema, Lichen nitidus, Lichen sclerosus, Erosive pustular dermatosis, Cutaneous sarcoidosis, Granuloma faciale, Granulomatous cheilitis. |
| Other Eruptions | Drug Rash, Acute generalized exanthematous pustulosis, Cutaneous lupus, Morphea/Scleroderma, Dermatitis herpetiformis, Dermatomyositis, Pityriasis rosea, Viral Exanthem, Erythema annulare centrifugum, Erythema multiforme, Erythema ab igne, Urticaria, Stasis Dermatitis, Photodermatitis, Hypersensitivity, Amyloidosis of skin, Keratosis pilaris, Prurigo nodularis, Inflicted skin lesions, Grover's disease, Confluent and reticulate papillomatosis, Keratoderma, Erythema nodosum, Zoon's balanitis, Xerosis, Lipodermatosclerosis, Retention hyperkeratosis, Lichenoid myxedema, Lymphomatoid papulosis, Skin striae, Sweet syndrome, Lichen striatus, Pruritic urticarial papules and plaques of pregnancy, Mastocytosis, Foreign body reaction of the skin, Fox-Fordyce disease, Pityriasis lichenoides, |

|  | Miliaria. |
|---|---|
| Neoplasms | SK/ISK, Melanocytic Nevus, Cyst, Verruca vulgaris, Lentigo, Dermatofibroma, Atypical Nevus, Scar Condition, Milia, Becker's nevus, Epidermal nevus, Nevus sebaceous, Skin Tag, Adnexal neoplasm, Pearly penile papules. Pilonidal cyst, Cutaneous neurofibroma, Comedone, Angiokeratoma of skin, Clavus, Porokeratosis, Fordyce spots, Benign neoplasm of nail apparatus, Angiofibroma, Accessory nipple, Glomus tumour of skin, Lichen Simplex Chronicus, Pyogenic granuloma, Chondrodermatitis nodularis, Knuckle pads, Actinic Keratosis, SCC/SCCIS, Basal Cell Carcinoma, Melanoma, Cutaneous metastasis, Cutaneous T Cell Lymphoma, Condyloma acuminatum |
| Blisters and Ulcers | Canker sore, Pyoderma Gangrenosum, Venous Stasis, Ulcer, Bullous Pemphigoid, Pemphigus vulgaris, Bullosis diabeticorum, SJS/TEN, Traumatic bulla |
| Nail disorders | Onychomycosis, Longitudinal melanonychia, Onychorrhexis, Onycholysis, Beau's lines<br>Onychopapilloma, Nail dystrophy due to trauma, Trachyonychia, Onychomadesis, Leukonychia, Onychocryptosis |
| Hair Disorders | Androgenetic Alopecia, Alopecia Areata. Telogen effluvium, Hirsutism, Lichen planopilaris, Dissecting cellulitis of scalp |
| Pigmentary Disorders | Acanthosis nigricans, Post-Inflammatory hyperpigmentation, Vitiligo, Melasma, Post-Inflammatory hypopigmentation, Pityriasis alba, Idiopathic guttate hypomelanosis, Erythema dyschromicum perstans, Hemosiderin pigmentation of skin |

| Vascular | Cutaneous capillary malformation, Hemangioma, Leukocytoclastic Vasculitis, Hematoma of skin, Livedo reticularis, Livedoid vasculopathy, Varicose veins of lower extremity |
|---|---|
| Others | Ecchymoses, Burn of skin, Erythromelalgia, Lymphedema, Ichthyosis, Flushing, Foreign body, Notalgia paresthetica |

**Supplementary Figure 1: UMAP projections before and after retraining the entire model.**
Points are colored by condition category. Refer to Figure 1C for legend.

| Original | With re-training |
|---|---|

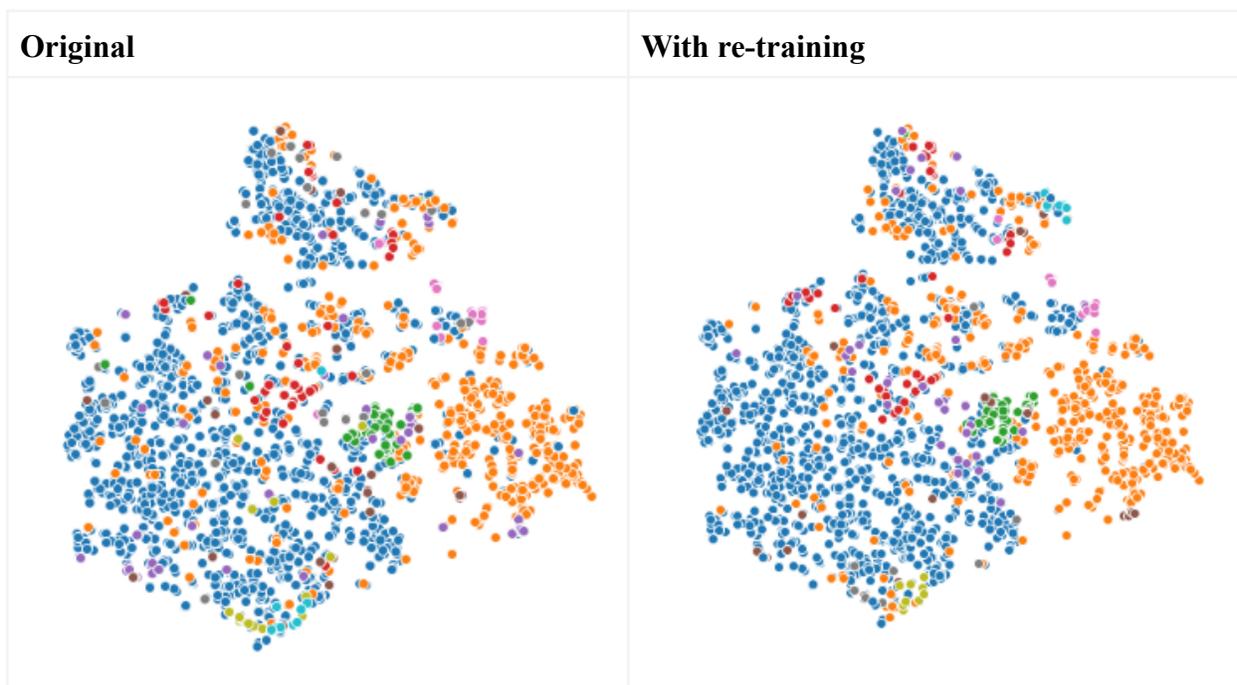

**Supplementary Figure 2: Examples of case images and respective saliency maps (computed via integrated gradients) before and after retraining.**

**A)**

| Original | After re-training |
|---|---|
| Healthy (0.337), Eczema (0.118), Tinea (0.093), Xerosis (0.093), Onychomycosis (0.052), Psoriasis (0.026), Onychocryptosis (0.022) | **Chillain (0.637)**, Healthy (0.118), Tinea (0.093), Xerosis (0.093), Onychomycosis (0.092), Onychocryptosis (0.022) |

**B)**

| Original | After re-training |
|---|---|
| **Allergic Contact Dermatitis (0.149)**, Eczema (0.103), Tinea (0.073), Lichen planus/lichenoid eruption (0.054), Herpes Zoster (0.052), Granuloma annulare (0.050), Psoriasis (0.041) | **Allergic Contact Dermatitis (0.749)**, Eczema (0.103), Tinea (0.073), Lichen planus/lichenoid eruption (0.075) |

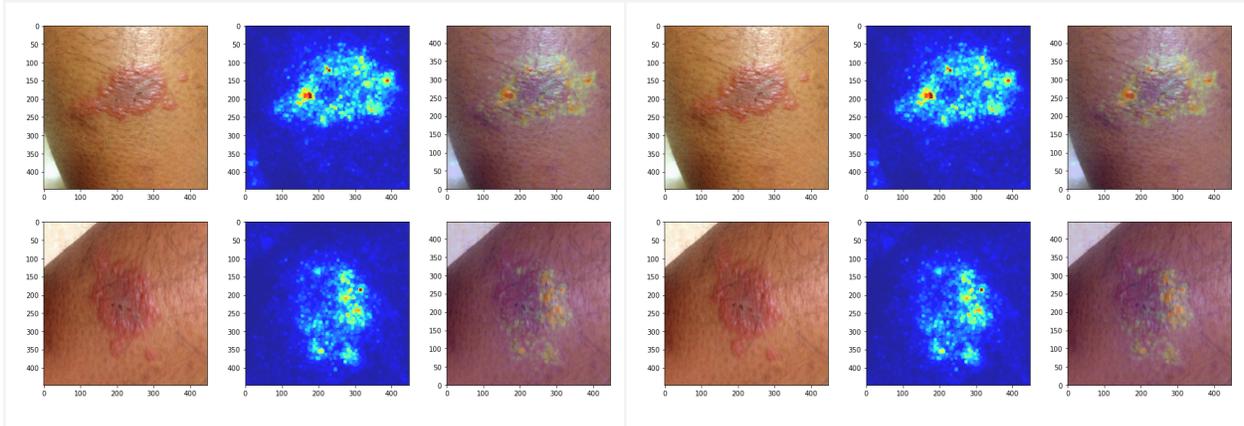

**Supplementary Figure 3: Visualizations similar to Figure 1C, but with 1 color per category for visual clarity.**

A. Model prediction accuracy

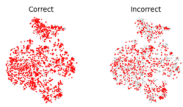

B. Condition category

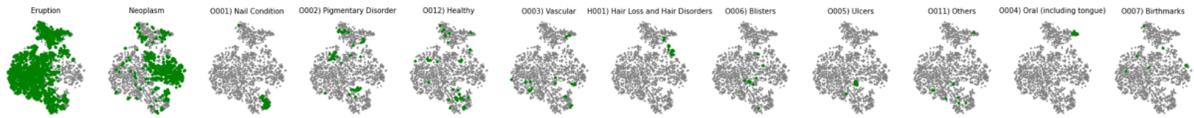

C. Lesion location (CLIN photos)

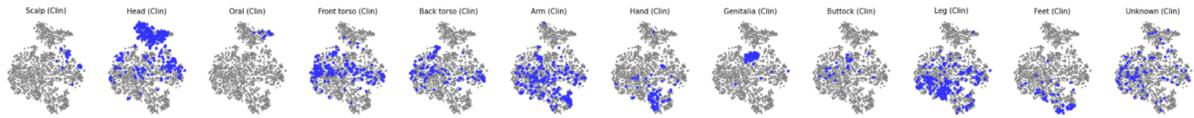

D. Lesion location (PAT photos)

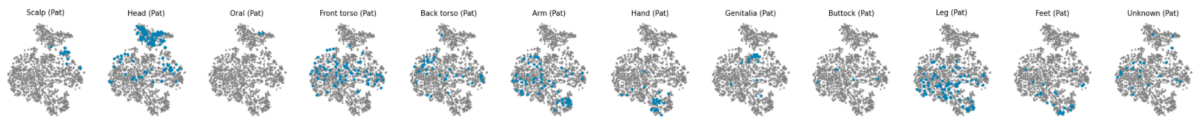